\documentclass[10pt]{article}
\usepackage{geometry}
\usepackage{graphicx} 
\usepackage{amsmath}
\usepackage{tikz}
\usetikzlibrary{arrows.meta}
\usepackage[ruled,vlined]{algorithm2e}
\usepackage{algpseudocode}
\usepackage[square, numbers]{natbib}
\usepackage{subcaption}
\usepackage{hyperref}

\title{Learning Temporally Consistent Turbulence Between Sparse Snapshots via Diffusion Models}
\author{M. Sardar$^{1}$ \and M. J. Zimo\'n$^{2, 3}$ \and S. Draycott$^{1}$ \and A. Revell$^{1}$ \and A. Skillen$^{1}$}
\date{
    $^{1}$ School of Engineering, The University of Manchester, Manchester, M13 9PL\\
    $^{2}$ School of Mathematics, The University of Manchester, Manchester, M13 9PL\\
    $^{3}$ IBM Research Europe, Daresbury, Warrington, WA4 4AD\\
}

\begin{document}

\maketitle

\section{Abstract}


We investigate the statistical accuracy of temporally interpolated spatiotemporal flow sequences between sparse, decorrelated snapshots of turbulent flow fields using conditional Denoising Diffusion Probabilistic Models (DDPMs). The developed method is presented as a proof-of-concept generative surrogate for reconstructing coherent turbulent dynamics between sparse snapshots, demonstrated on a 2D Kolmogorov Flow, and a 3D Kelvin-Helmholtz Instability (KHI). We analyse the generated flow sequences through the lens of statistical turbulence, examining the time-averaged turbulent kinetic energy spectra over generated sequences, and temporal decay of turbulent structures. For the non-stationary Kelvin-Helmholtz Instability, we assess the ability of the proposed method to capture evolving flow statistics across the most strongly time-varying flow regime. We additionally examine instantaneous fields and physically motivated metrics at key stages of the KHI flow evolution.

\section{Introduction}

Many engineering flows of interest require simulation of turbulence and its interaction with obstructions in some manner. In order to facilitate this, turbulence is often required as an inlet condition \cite{lund1998}. Many methods have been developed to generate realistic turbulence at the inlet to some region \cite{wu2016}. Typically the strong recycling method -- wherein the mid-plane of a periodic precursor simulation is taken as the inlet to the simulation of interest -- is preferred in Direct Numerical Simulations (DNS) of turbulent flows, however this approach may introduce repetitive flow patterns as artefacts to the main simulation \cite{jewkes2008}. Additionally, the storage requirements attached to DNS can hinder the acquisition rate of turbulent flow fields from precursor simulation. Thus, surrogates for inflow generation are an active area of study.

A related challenge arises in experimental measurement of turbulent flows, particularly in liquid metals, where acquisition rates may be as low as 1Hz \cite{vogt2018} which would be insufficient to resolve all characteristic turbulent time scales. Techniques such as ultrasound Doppler velocimetry or particle image velocimetry in high-speed flows often yield temporally sparse snapshots due to hardware limitations. Reconstructing the intermediate dynamics from sparse flow field observations would enable richer statistical analysis. 

Both problems share a common requirement: the need for statistically plausible turbulent dynamics between sparsely sampled flow states. 

Recent developments in Machine Learning (ML) have shown that temporal information present in sequences of turbulent flow fields can be exploited to produce synthetic spatiotemporal turbulence, using approaches designed for video generation. \citet{fukami2019d} demonstrated that an autoregressive Convolutional Neural Network (CNN) could be employed to generate sequences of turbulence, with some success. Additionally, they noted that their method was free of the spurious periodic artifacts which arise from the traditional recycling approach. \citet{yousif2022a} extended this approach using a physics-informed CNN combined with a model for time-series forecasting to generate sequences, showing improvements over the baseline method. Despite this, the ability of CNNs to reproduce high-wavenumber turbulent flow components was limited. Subsequent work in this area has focused on the use of Generative ML in the form of Generative Adversarial Networks (GANs). \citet{corsini2020} demonstrated that a standard GAN could learn to produce spatiotemporal turbulent flow data for synthetic inflow. \citet{kim2020} extended the GAN approach to include a Recurrent Neural Network (RNN) which learnt to generate noise which corresponded to successive timesteps, improving temporal coherence of generated sequences. They reported strong results for long rollout times, but noted limitations with regard to the quality of sequence generation when compared to the generation of instantaenous snapshots.  

Further developments in generative ML have shown that Denoising Diffusion Probabilistic Models (DDPMs) \cite{hoDenoisingDiffusionProbabilistic2020} can improve on GANs for image synthesis \cite{dhariwal2021}. Current state-of-the-art models for video generation are video DDPMs \cite{ho2022b}. 
In the context of turbulence, DDPMs have demonstrated their versatility through flexible conditioning approaches. \citet{gao2024} showcase this across super-resolution, unconditional spatiotemporal generation, and conditional generation using both gradient-based and concatenation-based conditioning. Their investigations into spatiotemporal generation focus on channel flow, which exhibits statistical stationarity, as a natural starting point for validating generative fidelity. Our work addresses a complementary setting: temporal infilling between sparsely sampled, decorrelated flow fields, evaluated both on stationary turbulence (Kolmogorov flow) and on non-stationary dynamics (KHI billow collapse) where the underlying statistics evolve throughout the generated interval. Other recent work has explored distinct aspects of diffusion-based turbulence generation. 
\citet{li2024} generate long trajectories of 1D Lagrangian turbulence and analyse the statistical properties of the generated tracer values. They extend this to infill between points in time for Lagrangian tracers of homogeneous isotropic turbulence as well as ocean surface drifters \cite{li2024f}. Their Lagrangian formulation benefits from trajectory continuity -- the particle's path provides implicit constraints on missing segments. In contrast, Eulerian field reconstruction between decorrelated snapshots lacks this trajectory constraint, and the model must generate spatial fields consistent with both instantaneous turbulence structure and temporal evolution. \citet{du2024} use a latent diffusion model with flexible gradient-based conditioning for a variety of flow reconstruction tasks, such as super-resolution from coarse representations and spatially sparse sensor data, as well as spatiotemporal field generation.
\citet{liu2024a} extend the method presented by \citet{du2024} for synthetic inflow generation, conditioning on upstream flow statistics and Reynolds number to produce inlet planes consistent with target flow parameters. \citet{kohl2024} demonstrate an autoregressive DDPM for sequential frame generation, propagating forward from a single initial condition. They note that temporal stability degrades over long rollouts -- a characteristic challenge of autoregressive approaches which endpoint-conditioned generation may circumvent by constraining both temporal boundaries. 

Unlike gap-filling or `inbetweening' approaches which exploit residual correlations between observed data and missing temporal segments, our method addresses the challenging task of generating plausible flow evolutions between temporally decorrelated states.  This capability is relevant for both the augmentation of sparse simulation data and enhancing temporally constrained experimental measurements. 


In this work, we develop and evaluate a conditional video DDPM for temporal infilling between sparse snapshots of turbulence, with potential applications in inflow generation, storage reduction for DNS simulations, and sparse-in-time experimental flow measurement techniques. We are not aware of prior work which systematically evaluates diffusion-based temporal infilling between sparse, decorrelated Eulerian turbulence snapshots, particularly in a non-stationary instability setting where flow statistics evolve rapidly.

We investigate the temporal evolution of a statistically stationary Kolmogorov Flow in 2D, and of 2D snapshots of a 3D Kelvin-Helmholtz Instability (KHI). We analyse our generated sequences through the lens of statistical turbulence, examining the time-averaged turbulent kinetic energy spectra over generated sequences, PDFs of quantities of interest, and how well POD modes are recovered in generated samples. Our contributions consist of:
\begin{itemize}
    \item Conditional temporal infilling between decorrelated snapshots of turbulence. 
    \item Evaluation on the statistically non-stationary and highly turbulent billow collapse phase of a Kelvin-Helmholtz Instability. 
    \item Comprehensive analysis through statistical turbulence metrics. 
\end{itemize}

\section{Methodology}
\label{sec:current work method}

Here we detail our approach to generate coherent turbulent trajectories on two canonical fluid flow problems: a statistically stationary Kolmogorov flow, and a non-stationary KHI. We employ a generative machine learning method, DDPMs, designed for video synthesis to generate coherent sequences of both flows.

\subsection{Case 1: Kolmogorov Flow Data}
\label{subsec:kol flow}

We generate 500 independent trajectories of the incompressible Navier-Stokes equations in 2D:

\begin{equation}
\begin{aligned}
\frac{\partial u_i}{\partial t} + u_j \frac{\partial u_i}{\partial x_j} & = - \frac{\partial p}{\partial x_i} + \frac{1}{\mathrm{Re}} \frac{\partial^2 u_i}{\partial x_j \partial x_j} + f_i, \\
\frac{\partial u_i}{\partial x_i} & = 0,
\end{aligned}
\label{eq:NSE}
\end{equation}

\noindent where $u_i$ is the dimensionless velocity field, $x_i$ is the spatial coordinate, $p$ is the dimensionless pressure, $\mathrm{Re}$ is the Reynolds number, $f_i = \sin{\left(10 \delta_{i2} x_1\right)}$ is a steady sinusoidal forcing term, and $\delta$ is the Kronecker delta. The domain is taken as a square of length $2 \pi$, with fully periodic boundaries. The equations are solved using a pseudo-spectral code with $256$ collocation points in each spatial direction and Fourier bases. We generate all trajectories at $\textrm{Re}=222$, allowing for turbulent dynamics. We take data only from the statistically stationary period of the flow, after the initial development transient. The flow is then separated into sequences of eight contiguous snapshots, which approximately corresponds to one decorrelation period. From our 500 independent trajectories, this yields a dataset of 2500 8-frame samples, of which we use 2000 for training, 450 for validation, and 50 for testing. 

\subsection{Case 2: Kelvin-Helmholtz Instability (KHI)}
\label{subsec: rbc}

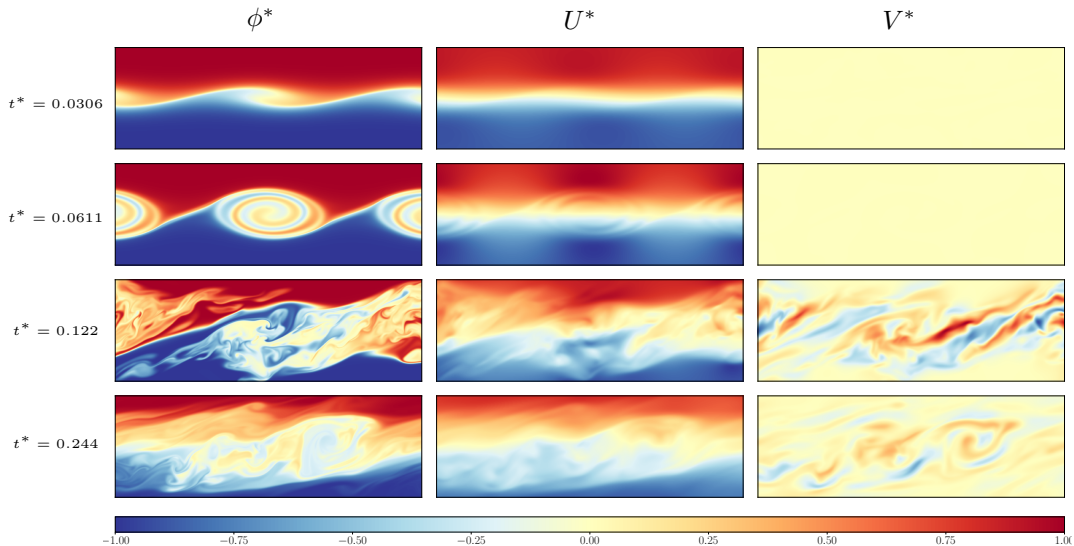
\begin{figure}
\centering
\begin{tikzpicture}
\node at (-3, -0.2) {$\phi^*$};
\node at (1.2, -0.2) {$U^*$};
\node at (5.4, -0.2) {$V^*$};
\node[font={\tiny}] at (-5.7, -1.3) {$t^* = 0.0306$};
\node[font={\tiny}] at (-5.7, -2.8) {$t^* = 0.0611$};
\node[font={\tiny}] at (-5.7, -4.3) {$t^* = 0.122$};
\node[font={\tiny}] at (-5.7, -5.8) {$t^* = 0.244$};
\node (kh) at (1.3, -3.5)
	{\includegraphics[width=0.8\textwidth, trim={0.5cm 5cm 0.5cm 2.8cm}, clip]{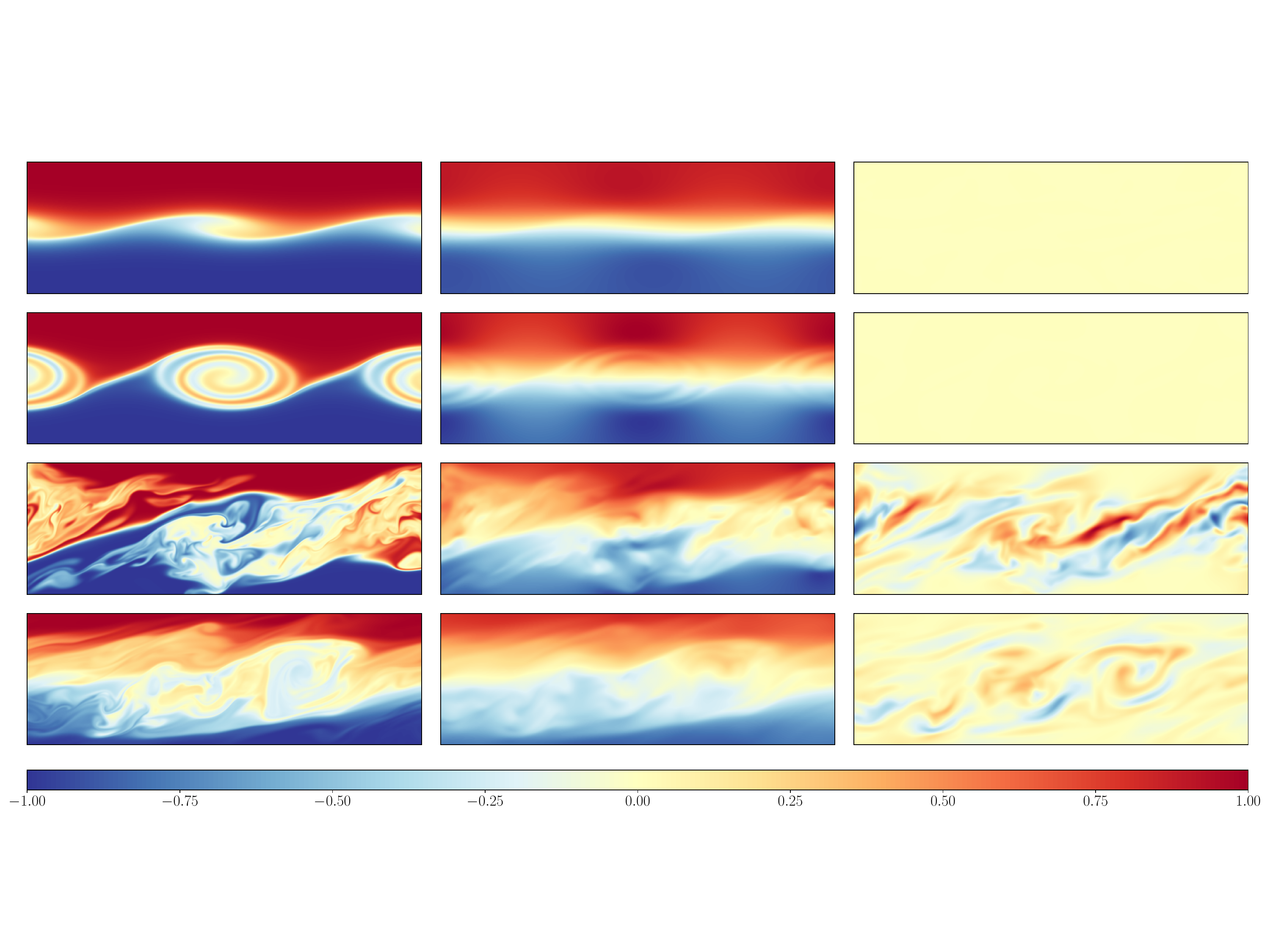}};
\end{tikzpicture}
\caption{Time evolution of a Kelvin-Helmholtz instability, generated via DNS. Spanwise component of velocity omitted. $\phi^*$ represents a dimensionless scalar, (i.e. $\theta^*)$ in \citet{smythLengthScalesTurbulence2000}. $U^*, V^*$ are the $x, y$ dimensionless velocity components, respectively. $t^*$ is dimensionless time, given by $t^* = t\frac{U_0}{L_x}$.}
\label{fig:kh_development}
\end{figure}

We investigate the evolution of a Kelvin-Helmholtz instability which is initiated by imposed inflectional instabilities in the shear region between two stably stratified flow. A buoyancy effect acts to dampen out turbulence generation due to background shear. Initially, due to the background shear, turbulence grows exponentially. At critical length scales, the buoyancy begins to first dampen the large scale structures, and eventually dampens out all turbulence, leaving a larger shear layer between the two flows, with a smaller density gradient across the interface. These stages are illustrated in Figure \ref{fig:kh_development}. 

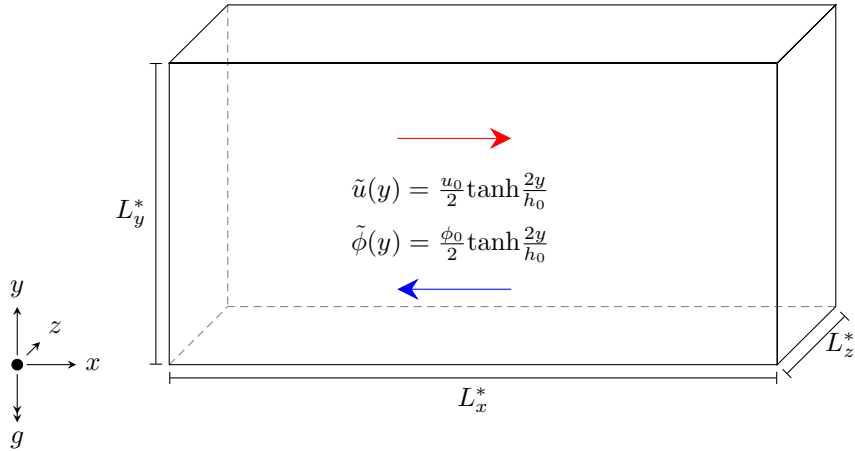
\begin{figure}[htpb!]
\centering
\begin{tikzpicture}[every edge quotes/.append style={auto, text=black}]
\pgfmathsetmacro{\cubex}{8}
  \pgfmathsetmacro{\cubey}{4}
  \pgfmathsetmacro{\cubez}{2}
  \draw [draw=black, every edge/.append style={draw=black, densely dashed, opacity=0.5}]
    (0,0,0) coordinate (o) -- ++(-\cubex,0,0) coordinate (a) -- ++(0,-\cubey,0) coordinate (b) edge coordinate [pos=1] (g) ++(0,0,-\cubez)  -- ++(\cubex,0,0) coordinate (c) -- cycle
    (o) -- ++(0,0,-\cubez) coordinate (d) -- ++(0,-\cubey,0) coordinate (e) edge (g) -- (c) -- cycle
    (o) -- (a) -- ++(0,0,-\cubez) coordinate (f) edge (g) -- (d) -- cycle;
  \path [every edge/.append style={draw=black, |-|}]
    (b) +(0,-5pt) coordinate (b1) edge node[below] {\(L^*_x\)} (b1 -| c)
    (b) +(-5pt,0) coordinate (b2) edge node[left] {\(L^*_y\)} (b2 |- a)
    (c) +(3.5pt,-3.5pt) coordinate (c2) edge node[right] {\(L^*_z\)} ([xshift=3.5pt,yshift=-3.5pt]e)
    ;
\node (base) at (-10, -4) {};
\node (x) at (-9, -4) {$x$};
\node (y) at (-10, -3) {$y$};
\node (g) at (-10, -5) {$g$};
\node (z) at (-9.5, -3.5) {$z$};
\draw[-stealth] (base) -- (x);
\draw[-stealth] (base) -- (y);
\draw[-stealth] (base) -- (z);
\draw[-{stealth}{stealth}] (base) -- (g);
\node (profile) at (-4.3, -1.7) {$\tilde{u}(y) = \frac{u_0}{2}$tanh$\frac{2y}{h_0}$};
\node (profile) at (-4.3, -2.4) {$\tilde{\phi}(y) = \frac{\phi_0}{2}$tanh$\frac{2y}{h_0}$};
\draw[-{Stealth[scale=4, length=2, width=2]}, red] (-5, -1) -- (-3.5, -1);
\draw[{Stealth[scale=4, length=2, width=2]}-, blue] (-5, -3) -- (-3.5, -3);
\draw[fill, black] (base) circle [radius=2pt];
\end{tikzpicture}
\caption{Simulation domain and intial flow conditions. Gravity vector depicted as $g$. Initial velocity profile and temperature profile given as a function of the velocity and length scales in the shear layer. $L^*_x = 6.8$, $L^*_y=13.65$, $L^*_z =1.6$, are the dimensionless lengths of the domain bounding box as a ratio to $h_0$. The top and bottom of the domain, $(y_{max}, y_{min})$, have zero gradient boundary conditions for $U, V, \phi$. The left, right, front, and back boundaries at $(x_{min}, x_{max}),\ (z_{min}, z_{max})$ are periodic.}
\label{fig:domain setup}
\end{figure}

Data from this case study is obtained via DNS, using an 8th order explicit finite difference scheme, with key dimensionless parameters listed in table \ref{table:1}. The Reynolds number, $Re_0$, Richardson number $Ri_0$, and Prandtl number $Pr$ are defined as,

\begin{equation}
Re_0 \equiv \dfrac{u_0 h_0}{\nu}, \qquad Ri_0 \equiv \dfrac{g\phi_0 h_0}{u^2_0}, \qquad Pr \equiv \dfrac{\nu}{\kappa_m}
\end{equation}

where $u_0$ represents the initial change in velocity across the shear layer, $u_0 = u_{max} - u_{min}$. The initial thickness of the shear layer, $h_0$, is prescribed to initialise the simulation. The initial temperature difference across the shear layer is $\phi_0$, $\nu$ is the fluid viscosity, and $\kappa_m$ is thermal diffusivity. The bulk Richardson number, $\textrm{Ri}_0$, describes the relationship between shear and stratification due to buoyancy. The bulk Richardson number is gradually driven towards a limit of $\frac{1}{4}$, where the flow stabilises (after the instability has decayed) \cite{smythLengthScalesTurbulence2000}. Figure \ref{fig:domain setup} illustrates the domain initialisation and axes used throughout.
\begin{table}[h]
\begin{center}
\begin{tabular}{c || c}
Parameter & Value \\
\hline
$\textrm{Re}_0$ & $1354$ \\
$\textrm{Pr}$ & $7$ \\
$\textrm{Ri}_0$ & $0.08$
\end{tabular}
\end{center}
\caption{Dimensionless parameters and their values. $\textrm{Re}_0$ is the Initial Reynolds number, $\textrm{Ri}_0$ is the initial bulk Richardson number, and $\textrm{Pr}$ is the Prandtl number.}
\label{table:1}
\end{table}

In total, six separate realisations of the flow are used (i.e. starting from different random seeds). Four are used for training, one for validation, and one is reserved for testing. The turbulent portion of the flow is extracted as a spatiotemporal subset; 32 frames are extracted from the point of billow collapse to the establishment of a new shear layer. These are separated further into four sequences of eight frames, taking 50 2D slices along the spanwise direction at each timestep. These are then postprocessed to a smaller spatial resolution, to filter high-wavenumber turbulence and focus on capturing temporal evolutions. 

\subsection{Diffusion Model}

Here we present a brief outline of the continuous-time DDPM formulation as presented in \citet{ho2022a}, and used in their video DDPM approach \cite{ho2022b}. These present an improvement over the initial DDPMs developed in \citet{hoDenoisingDiffusionProbabilistic2020}. 

\begin{figure}[ht!]
\centering
\begin{subfigure}{\textwidth}
\begin{tikzpicture}
     \node (1) at (-1.6, 0)
     {\includegraphics[width=0.8\textwidth, trim=0 0 0 0, clip]{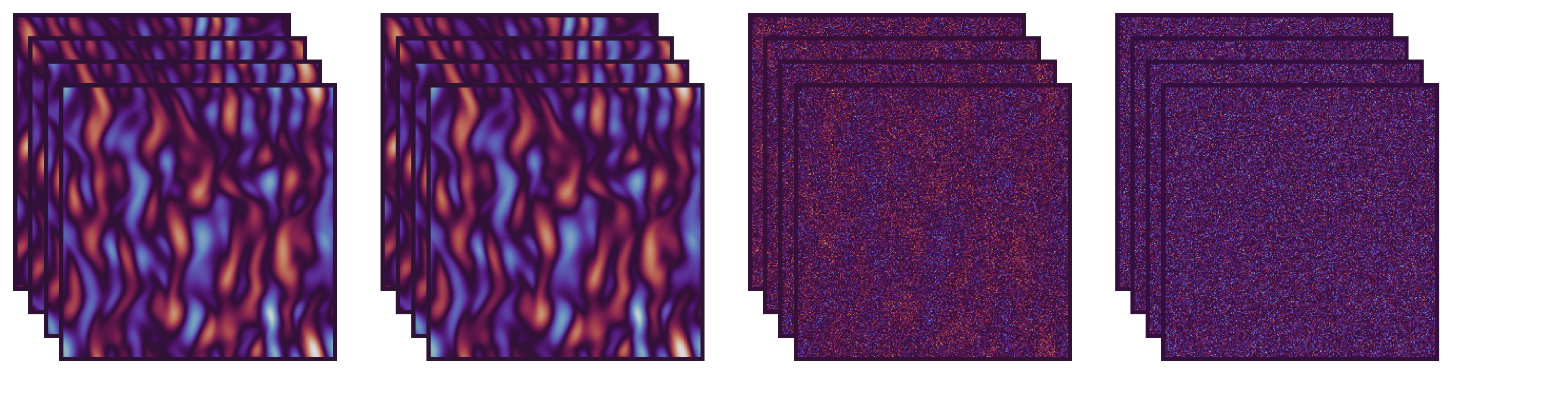}};
     \node (2) at (5.6, 0)
     {\includegraphics[width=0.2\textwidth, trim=0 0 0 0, clip]{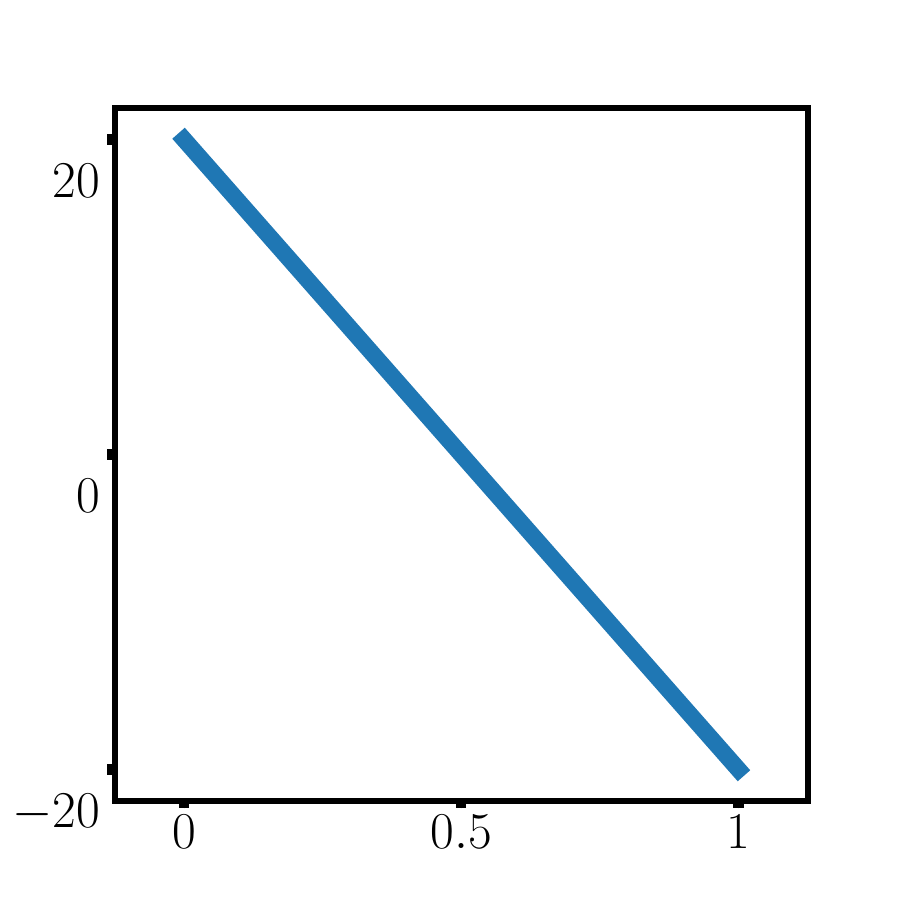}};
     \node (lambda) at (4.1, -0.12){\small $\lambda$};
     \node (lambda) at (5.7, -1.65){\small $\lambda^{*}$};
\end{tikzpicture}
\end{subfigure}

\begin{subfigure}{\textwidth}
\begin{tikzpicture}
     \node (1) at (-1.6, 0)
     {\includegraphics[width=0.8\textwidth, trim=0 0 0 0, clip]{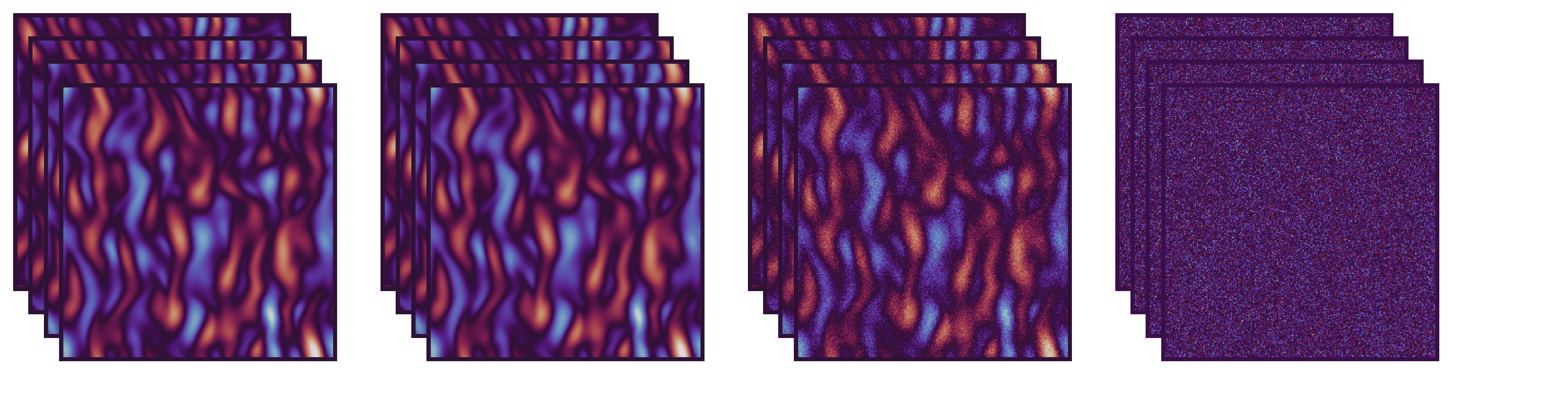}};
     \node (2) at (5.6, 0)
     {\includegraphics[width=0.2\textwidth, trim=0 0 0 0, clip]{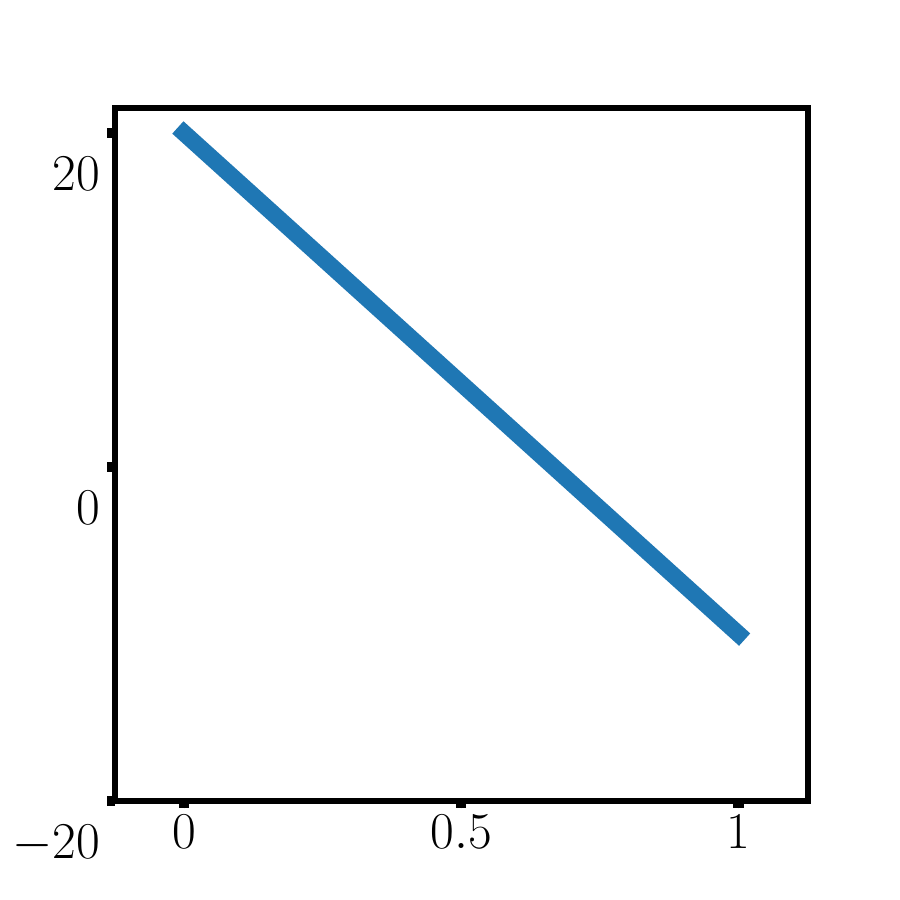}};
     \node (lambda) at (4.1, -0.12){\small $\lambda$};
     \node (lambda) at (5.7, -1.65){\small $\lambda^{*}$};
\end{tikzpicture}
\end{subfigure}
\caption{A comparison of two choices of $\{\lambda_{min}, \lambda_{max}\}$ in terms of their impact on a sample from the Kolmogorov flow dataset.}
\label{fig:noising schedules}
\end{figure}
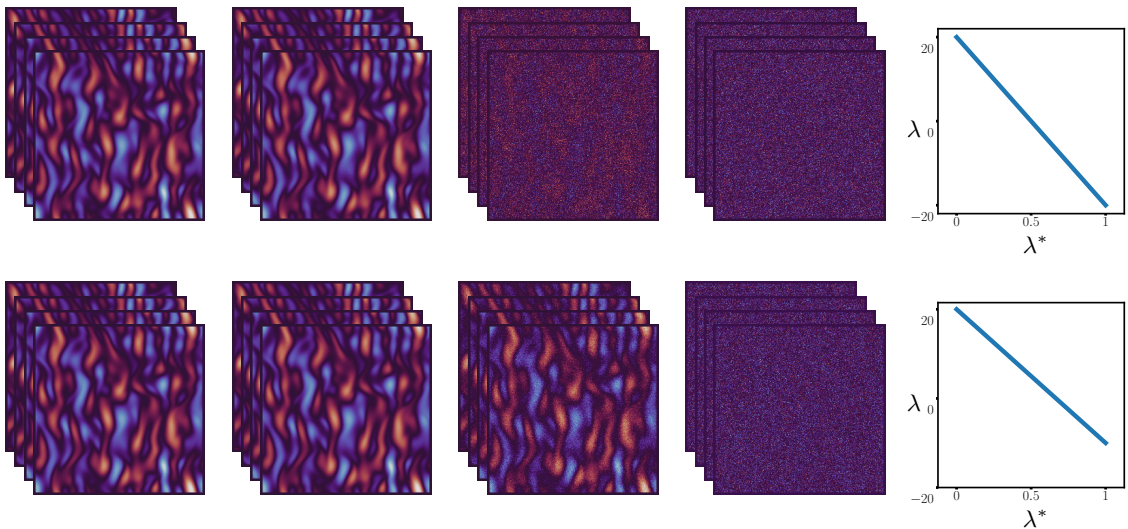

DDPMs are a novel class of generative method which rely on learning the reverse process induced by a forward diffusion process. The forward process is described by a Markov chain of Gaussian transition kernels, which have the property that any step in the forward process can be `jumped' to through the cumulative effect of the Gaussian transitions (which by definition is itself a Gaussian). The forward process can be understood as an information degradation process. Typically, the process is defined discretely in the form of a series of variances which define a piecewise-smooth transition from data to noise. By learning the reverse process, a diffusion model is able to generate samples of data from pure noise. The continuous-time DDPM formulation presented in \citet{ho2022} allows for a continuous parameterisation of the forward process by conditioning directly on $\lambda$, the log Signal-to-Noise Ratio. The functional form of $\lambda$ varies with various noising strategies; we employ a linear relationship.  Figure \ref{fig:noising schedules} presents two choices of $\{\lambda_{min}, \lambda_{max}\}$ alongside samples noised to different $\lambda^{*}$ between them, one in which information is destroyed quickly, and one in which the information is destroyed more gradually. Here, $\lambda^{*}$ is used to denote $\lambda$ normalised between 0 and 1. Note that the choice of $\{\lambda_{min}, \lambda_{max}\}$ is a hyperparameter, which as per \citet{ho2022a}, we fix to $\{-20, 20\}$ for all experiments.

We use a video DDPM \cite{ho2022b} to generate new realisations of the trajectory between decorrelated states of a velocity field for Kolmogorov flow, and a temperature field for the KHI. A UNet \cite{ronnebergerUNetConvolutionalNetworks2015} CNN with 3D convolutions is used to parameterise the noise in noisy samples. During training, we show the DDPM a noised ground-truth DNS sequence, with $n_{cond}$ frames of un-noised data passed in as conditioning information. We compute $n_{cond}$ as a mask.

\begin{algorithm}
\SetAlgoLined
\KwIn{$y_0$ (input tensor), $n\_cond\_frames$ (number of condition frames)}
\KwOut{mask (conditional mask tensor)}

$L \gets \text{shape}(y_0)[2] - 1$ \tcp*{Last index of the third dimension}

\tcp{Calculate indices for condition frames}
$cond\_indices \gets \emptyset$\;
\For{$j \gets 0$ \KwTo $n\_cond\_frames - 1$}{
    $index \gets \lfloor j \cdot L / (n\_cond\_frames - 1) \rfloor$\;
    $cond\_indices \gets cond\_indices \cup \{index\}$\;
}

Initialize $mask$ as an empty tensor\;

\For{$i \gets 0$ \KwTo $L$}{
    \eIf{$i \in cond\_indices$}{
        Append ones to $mask$\;
    }{
        Append zeros to $mask$\;
    }
}

\Return{$mask$}
\caption{Conditional Mask Generation}
\label{alg: mask generation}
\end{algorithm}

Ground truth samples of eight-frame snapshots are passed through the forward noising process (Figure \ref{fig:noising schedules}), and concatenated to $n$ conditioning frames. The sequence of conditioning frames is preserved by setting the conditioning frame values to zero in those frames not used for conditioning, which we present in Algorithm \ref{alg: mask generation}. The noise-parameterising UNet is trained to predict the added noise to the original video. For classifier-free diffusion guidance, we follow \cite{ho2022} and set $p_{uncond}=0.2$.

At inference time, the UNet is used as one component of an iterative scheme for generation. Generation is carried out by discretising the functional form of the log(SNR) using the limits of $\{\lambda_{min}, \lambda_{max}\}$ chosen during training, and iteratively traverses the associated reverse Markov chain to generate a new sample from the original data distribution, conditional on the provided conditioning frames. The reverse process is defined as 
\newline $q\left(\boldsymbol{z}_{\lambda^{\prime}} | \boldsymbol{z}_{\lambda}, \boldsymbol{x}\right) = \mathcal{N}\left(\tilde{\boldsymbol{\mu}}_{\lambda^{\prime} | \lambda}\left(\boldsymbol{z}_{\lambda}, \boldsymbol{x}\right), \tilde{\sigma}^2_{\lambda^{\prime} | \lambda}\boldsymbol{I}\right)$, with terms defined in Eq. \ref{eq: cddpm mean variance}\cite{ho2022}:

\begin{equation}
\boldsymbol{\tilde{\mu}}_{\lambda^{\prime} | \lambda} = e^{\lambda - \lambda^{\prime}} \left(\dfrac{\alpha_{\lambda^{\prime}}}{\alpha_{\lambda}}\right) \boldsymbol{z}_{\lambda} + (1 - e^{\lambda - \lambda^{\prime}})\alpha_{\lambda^{\prime}}\boldsymbol{x}, \quad \tilde{\sigma}^2_{\lambda^{\prime} | \lambda} = (1 - e^{\lambda - \lambda^{\prime}})\sigma^2_{\lambda^{\prime}},
\label{eq: cddpm mean variance}
\end{equation}

where $\boldsymbol{\tilde{\mu}}_{\lambda^{\prime} | \lambda}$ is the reverse process mean at some $\lambda^{\prime} < \lambda$ given the mean at $\lambda$, $\alpha_{\lambda}=\sqrt{1/(1+e^{-\lambda})}$, $z_{\lambda}$ represents a sample from the target distribution noised to $\lambda$, $\boldsymbol{x}$ represents the original noise-free video, $\sigma^2_{\lambda}$ is the forward process variance at $\lambda$, and $\tilde{\sigma^2_{\lambda^{\prime} | \lambda}}$ is the reverse process variance at some $\lambda^{\prime} < \lambda$.

As our process is generative, $\boldsymbol{x}$ is instead parameterised by the trained UNet, such that $\boldsymbol{x}_{\theta}(\boldsymbol{z}_{\lambda}) = (\boldsymbol{z}_{\lambda} - \sigma_{\lambda}\boldsymbol{\epsilon}_{\theta}(\boldsymbol{z}_{\lambda}, \boldsymbol{c}, \lambda))/\alpha_{\lambda}$, where $\boldsymbol{\epsilon}_{\theta}(\boldsymbol{z}_{\lambda}, \boldsymbol{c}, \lambda)$ is our trained UNet, given an input noisy image, conditioning information in the form of $n$ conditioning frames, and $\lambda$. 

As mentioned above, the UNet is trained partially on the unconditional generation task. This is then used during inference, with a guidance strength parameter, $w$, to obtain an interpolated value for the noise prediction during inference, $\tilde{\epsilon}_{\theta} = (1 + w)\epsilon_{\theta}(\boldsymbol{z}_{\lambda}, \boldsymbol{c}, \lambda) - w\epsilon_{\theta}(\boldsymbol{z}_{\lambda},\lambda)$. The values of $w$ and $v$ are obtained via hyperparameter optimisation, with values suggested in \citet{ho2022} as starting points.

The initial generative process starts at $p_{\theta}\left(\boldsymbol{z}_{\lambda_{min}}\right)=\mathcal{N}\left(\boldsymbol{0}, \boldsymbol{I}\right)$. The transitions defined in Eq. \ref{eq: cddpm mean variance} then become Eq. \ref{eq: cddpm parameterised transitions}:

\begin{equation}
p_{\theta}\left(\boldsymbol{z}_{\lambda^{\prime}} | \boldsymbol{z}_{\lambda}\right) = \mathcal{N}\left(\boldsymbol{\tilde{\mu}}_{\lambda^{\prime} | \lambda}\left(\boldsymbol{z}_{\lambda}, \boldsymbol{x}_{\theta}\right),\left(\tilde{\sigma}^2_{\lambda^{\prime} | \lambda}\right)^{1 - v}\left(\sigma^2_{\lambda | \lambda^{\prime}}\right)^{v}\right)
\label{eq: cddpm parameterised transitions}
\end{equation}

where, for classifier-free guidance during inference, $v$ gives an interpolation strength between the variance for unconditional and conditional generation. 

\section{Results}
\label{sec:results}

\subsection{Case 1: Synthetic Trajectories of a Kolmogorov Flow}

We present the results of a video DDPM trained to perform a temporal interpolation between frames of a Kolmogorov Flow. These frames consist of eight contiguous snapshots, each of $256\times256$ grid points, with the first and last snapshot being approximately temporally decorrelated. We evaluate our approach using a range of metrics, including the autocorrelation function, structure functions, energy spectra, probability distributions of vorticity, and instantaneous statistics. We note that our aim is not to recover identical snapshots to the corresponding DNS, but rather to generate synthetic, statistically equivalent realisations of posssible trajectories between states of a turbulent flow field. 

\begin{figure}
\centering
\includegraphics[width=0.8\textwidth]{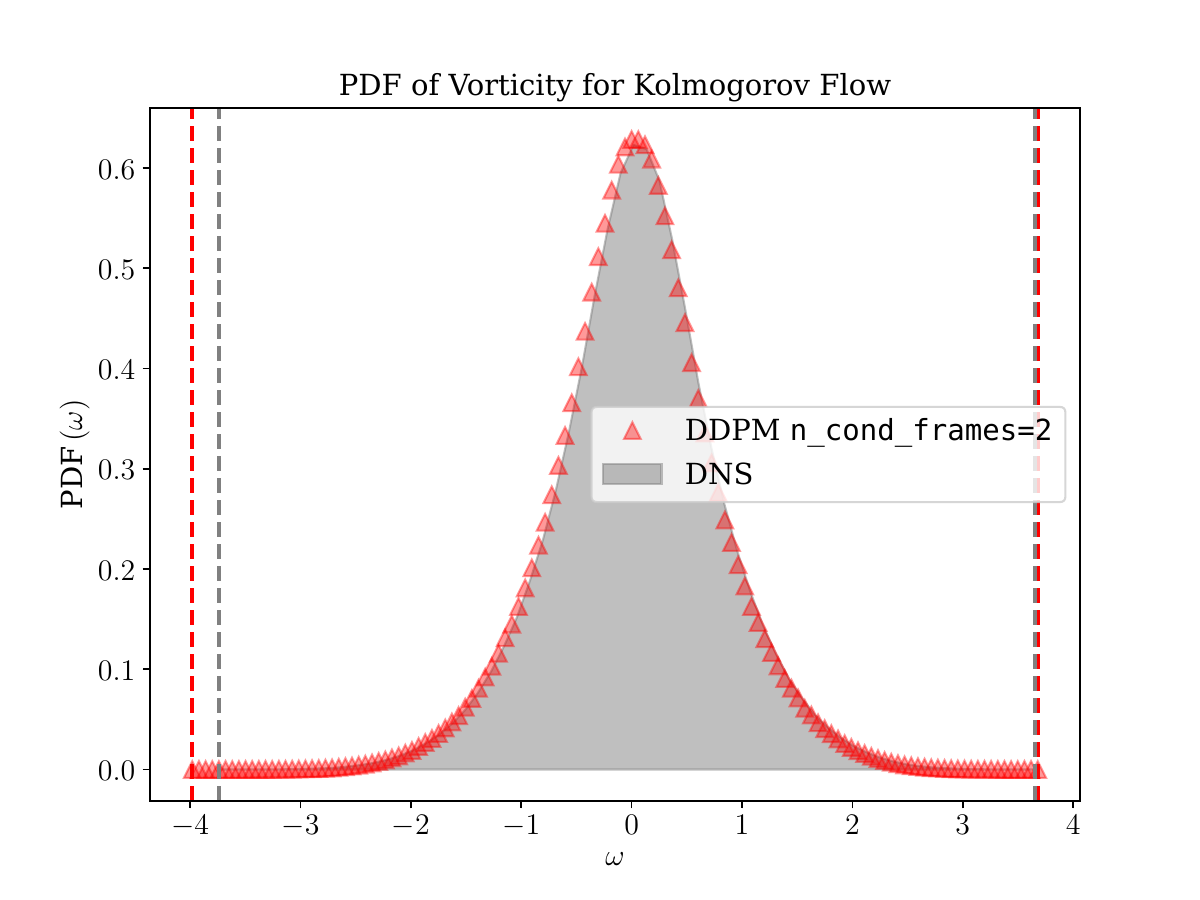}
\caption{A comparison of the PDFs of vorticity for: the reference DNS sequences, and DDPM-generated outputs with different numbers of conditioning frames. Upper and lower limits are indicated for each PDF in dashed lines.}
\label{fig:pdf kol}
\end{figure}

A chaotic turbulent flow may be characteristed statistically in terms of its Probability Distribution Function (PDF). In Figure \ref{fig:pdf kol}, we present PDFs of the vorticity across 50 generated sequences, alongside the PDF of the corresponding DNS snapshots.

\begin{figure}
\centering
\includegraphics[width=0.9\textwidth]{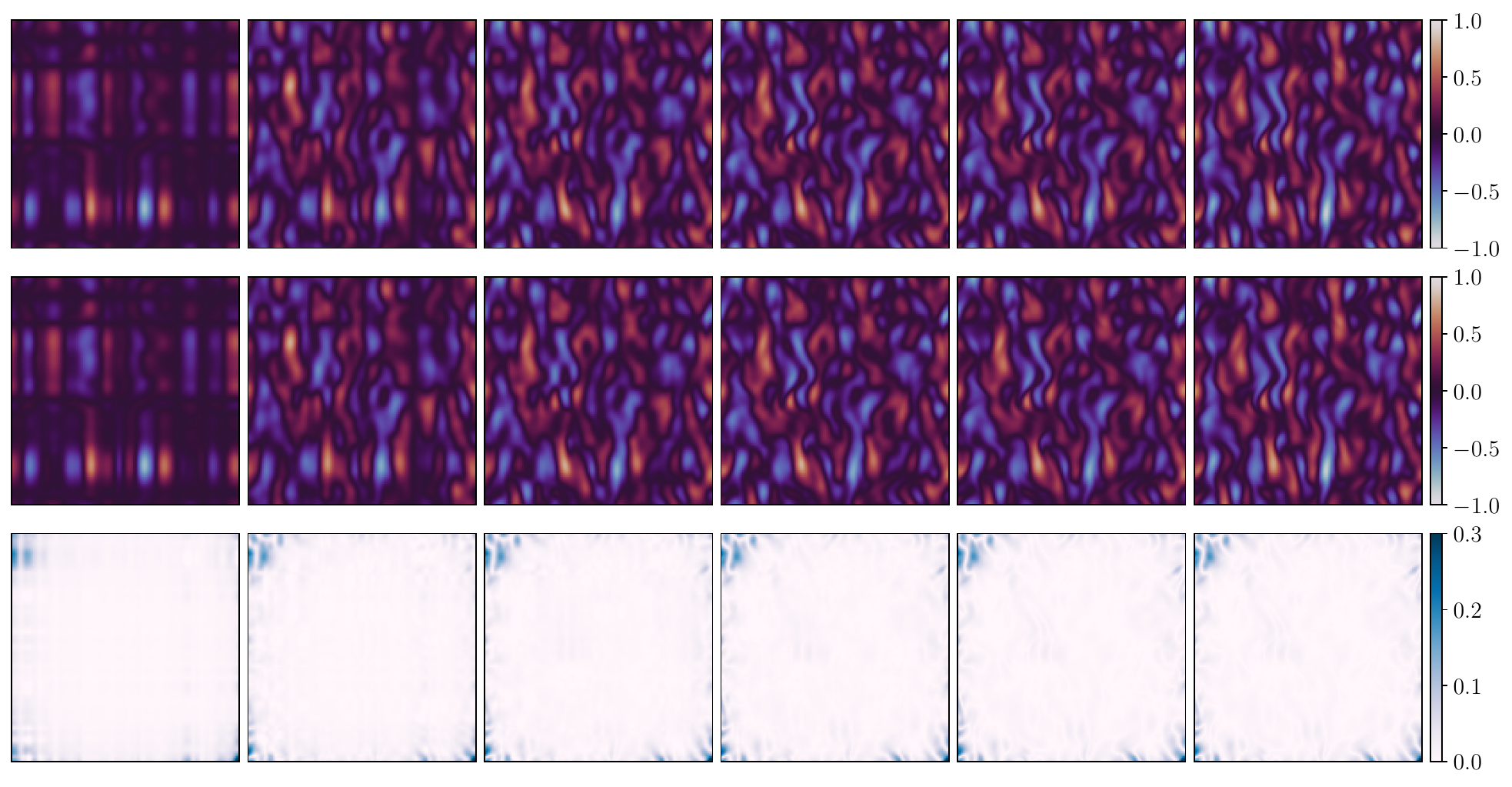}
\caption{A comparison of POD mode-based reconstructions of a sample DNS sequence from the test set (top), the corresponding DDPM-generated sequence (middle), and the error field between them (bottom) shown for $u_x$ of the mid-sequence snapshot. From left to right, the POD modes used for reconstruction are: $1, 5, 10, 20, 50$, with the final plot showing the actual field.}
\label{fig:pod kol}
\end{figure}

The instantaneous recovery of fields can be examined through a reconstruction of the Proper Orthogonal Decomposition (POD) modes of snapshots. POD modes identify coherent structures, ranked by their energy content; each POD mode can be thought of as a combination of fourier modes. In Figure \ref{fig:pod kol}, we present the reconstructions of $u_x$ for the mid-sequence snapshot of a representative sample, with the ground truth data and DDPM derived sequence. The mid-sequence snapshot is selected as the generated frame `furthest' from the conditioning information. Qualitatively, our model recovers the primary POD modes well, and the higher POD modes very well. 

\begin{figure}
\centering
\includegraphics[width=0.8\textwidth]{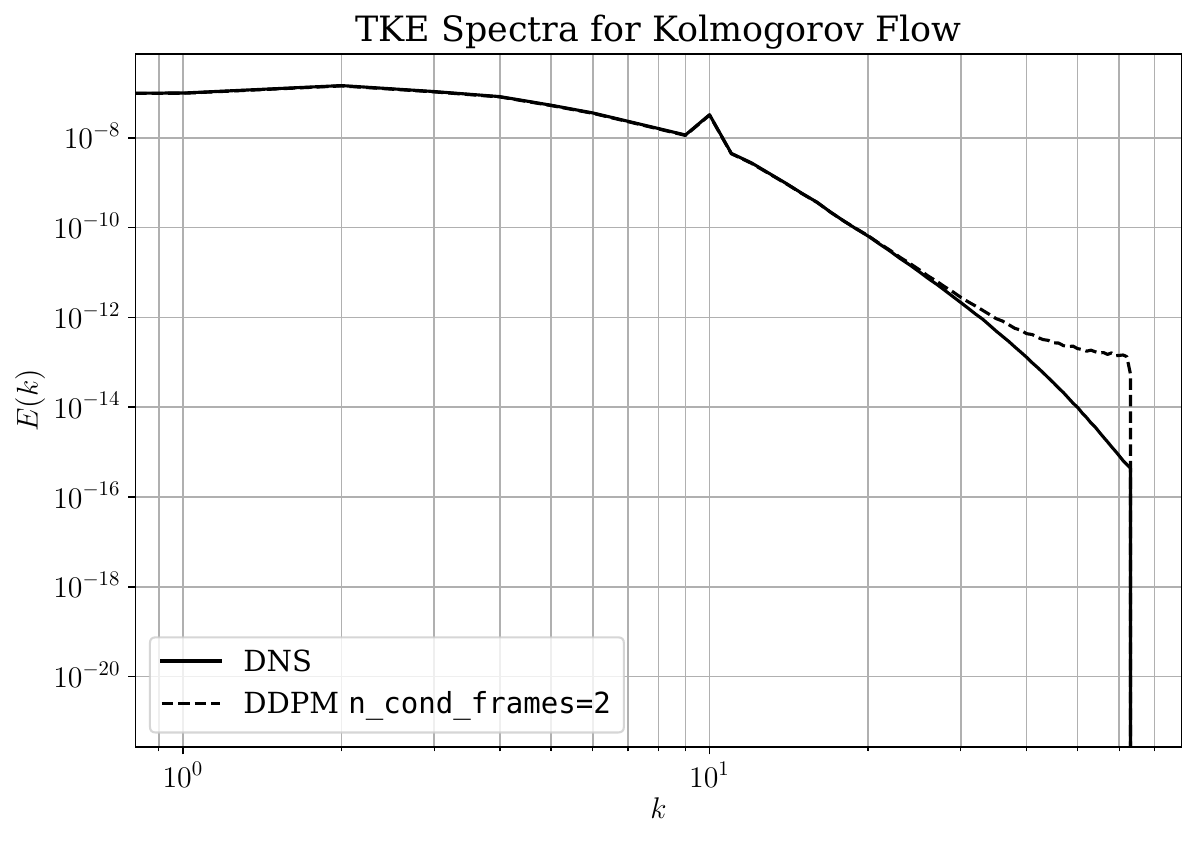}
\caption{A comparison of the TKE spectra of the ground truth DNS sequences and the DDPM generated sequences.}
\label{fig:s1 spectra}
\end{figure}

In order to investigate the recovery of information at all length scales of turbulence, we must examine the turbulent kinetic energy spectra. Figure \ref{fig:s1 spectra} shows that through the temporal interpolation, the generated sequences are statistically coherent with the underlying DNS up to a given wavenumber, as expected from our prior work in super-resolution of a Kolmogorov flow, where it was found that super-resolution using a DDPM yielded generated samples with the correct TKE spectrum up to a given wavenumber limit \cite{sardar2024}.


\subsection{Case 2: Synthetic Trajectories of a Kelvin-Helmholtz Instability}
\label{subsec:ffdm}

Here we present the results of a spatiotemporal sequence generation carried out using a video DDPM, as before, on a more complex case. A Kelvin-Helmholtz Instability contains highly complex turbulent behaviour, with the flow moving through different regimes as billows form, couple, and collapse. This test case allows us to investigate the ability of video DDPMs to interpolate between statistically non-stationary fields, and we observe that they are able generate new realisations of the transition between different stages of the flow. 

We present the results of a video DDPM trained to perform a temporal interpolation between frames of a Kelvin-Helmholtz Instability. These frames consist of eight contiguous snapshots, each of $72\times208$ grid points, with the first and last snapshot being approximately temporally decorrelated. As the statistics are evolving in time, each sequence is assigned a label based on where it is in the flow's temporal evolution -- for example, the billow collapse phase is given the label $1$. 
We evaluate our approach using a range of physically motivated metrics. We note again that our aim is not to recover identical snapshots to the corresponding DNS, but rather to generate synthetic, statistically equivalent realisations of posssible trajectories between states of a turbulent flow field. We analyse the results from the billow collapse phase, as this is where the statistics change most rapidly in time.

\begin{figure}
\centering\captionsetup{width=\linewidth}
\begin{subfigure}{\textwidth}
\includegraphics[width=\textwidth, trim=0 15 0 0, clip]{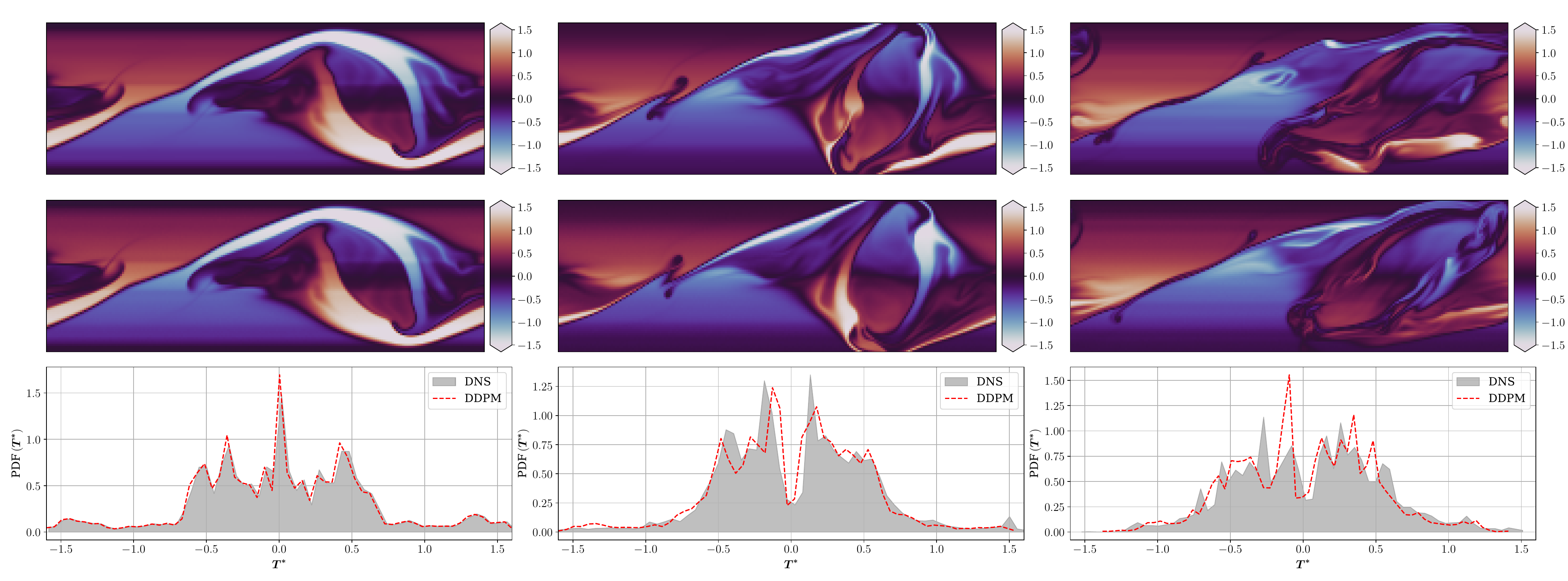}
  \setlength{\belowcaptionskip}{-2pt}
  \caption{Frames $\{0, 1, 2\}$}
  \label{fig: kh pdfs 012}
\end{subfigure}

\begin{subfigure}{\textwidth}
\includegraphics[width=\textwidth, trim=0 15 0 0, clip]{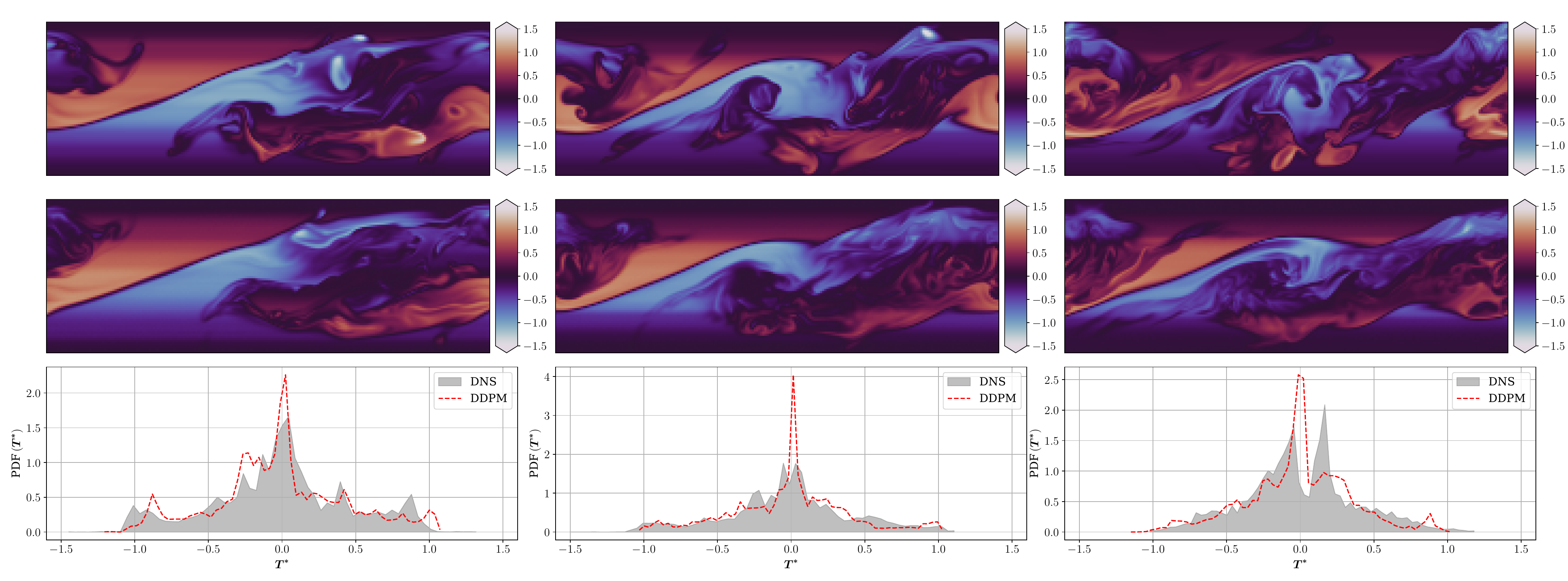}
  \setlength{\belowcaptionskip}{-2pt}
  \caption{Frames $\{3, 4, 5\}$}
  \label{fig: kh pdfs 345}
\end{subfigure}

\begin{subfigure}{0.66\textwidth}
\includegraphics[width=\textwidth, trim=0 15 0 0, clip]{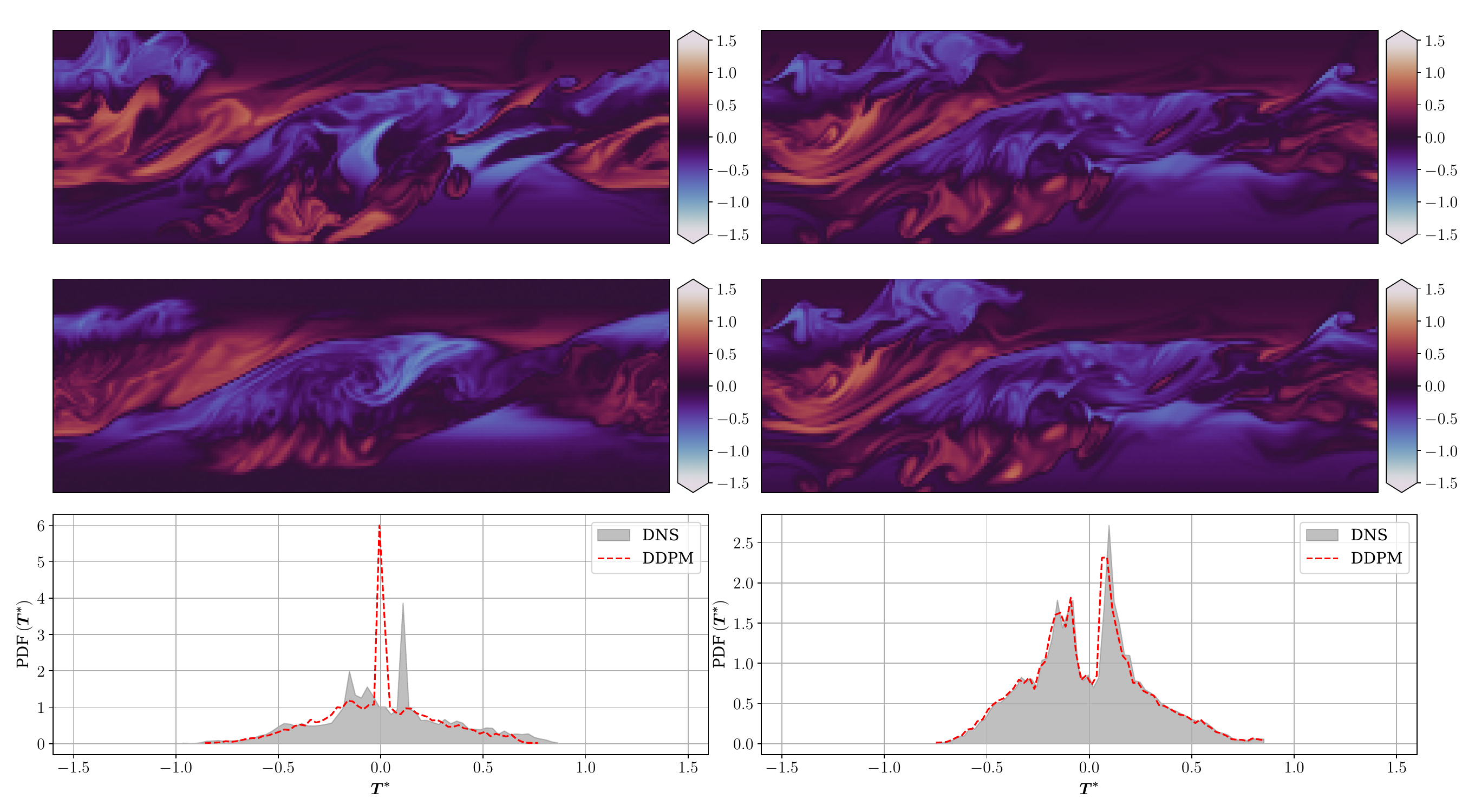}
  \setlength{\belowcaptionskip}{-2pt}
  \caption{Frames $\{6, 7\}$}
  \label{fig: kh pdfs 67}
\end{subfigure}
\setlength{\belowcaptionskip}{-1pt}
\caption{Representative sequence frames with DNS data (top), DDPM-generated data (middle), and $\textrm{PDF}\left(T^{\prime}\right)$ (bottom).}
\label{fig: kh pdfs}
\end{figure}

As was carried out for the Kolmogorov flow, we may characterise certain quantities as random variables and analyse the statistical recovery of these quantities through our approach. In this case, we take the velocity fluctuations, $T^\prime = T - \langle T\rangle$, where $\langle \cdot \rangle$ is an averaging operator, defined here as averaging across spanwise locations due to the changing statistics in time. Figure \ref{fig: kh pdfs} shows frames from a representative sequence: frames $\{0, 7\}$ are given as conditioning information, with the remainder being temporally infilled by the DDPM. The model captures the statistics well, with some slight deviations from the DNS PDF observable in certain frames. 

\begin{figure}
\centering
\includegraphics[width=\textwidth, trim=0 160 0 160, clip]{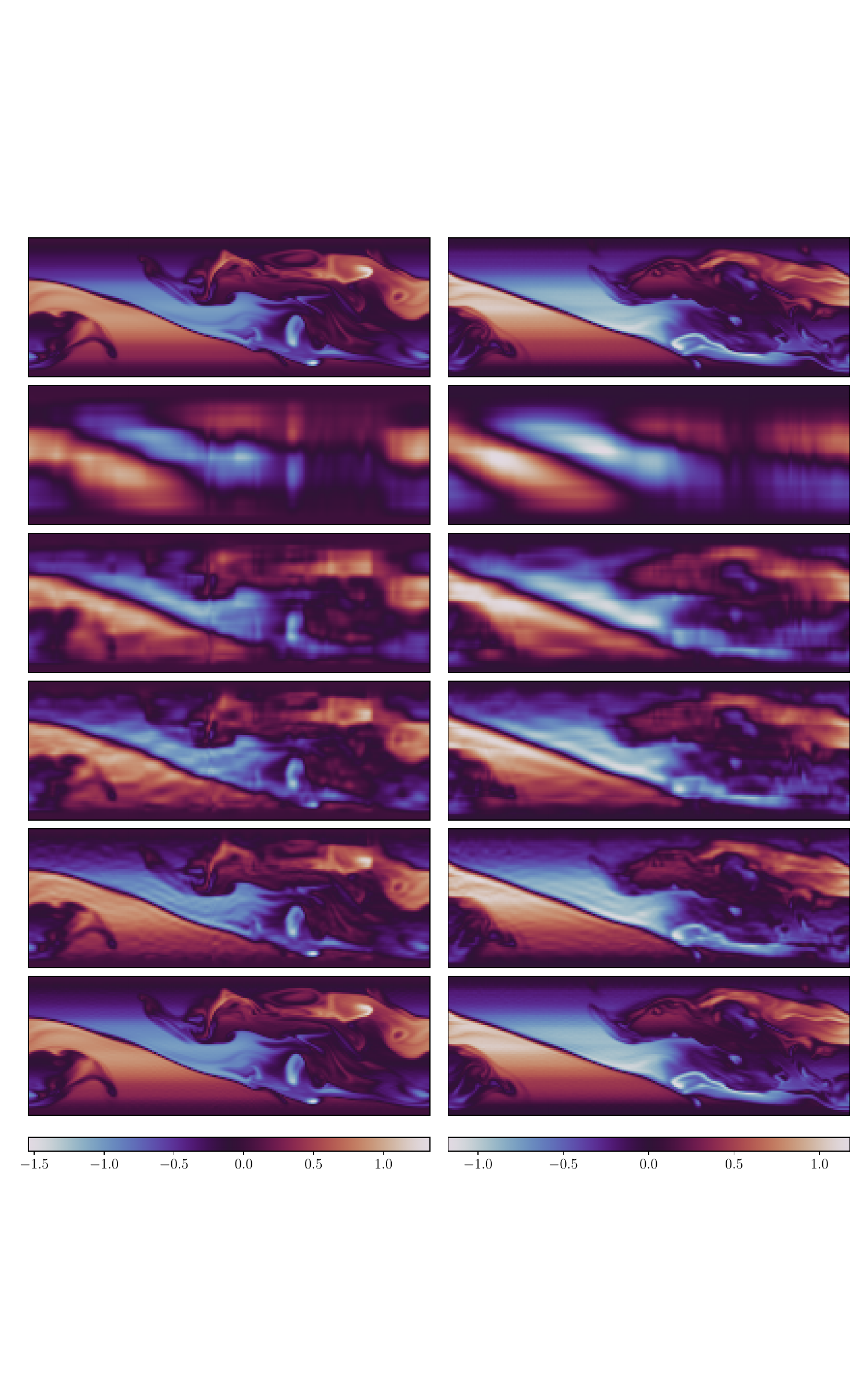}
\caption{A comparison of POD mode-based reconstructions of a sample DNS sequence (left) and a DDPM-generated sequence (right), shown for $T^{\prime}$ of the mid-sequence snapshot (frame 3). From top to bottom, the POD modes used for reconstruction are: $1, 5, 10, 20, 50$, with the first row showing the actual field.}
\label{fig: kh pod modes}
\end{figure}

POD modes are analysed for snapshots of the Kelvin-Helmholtz instability to verify whether bulk structures are recovered well. Figure \ref{fig: kh pod modes} shows that while coherent structures are recovered, their location is not necessarily the same as in the ground truth data. This is expected from DDPM outputs, where matching statistics is prioritised over the exact reconstruction of samples; this leads to higher sample diversity. 

\begin{figure}
\centering
\includegraphics[width=\textwidth]{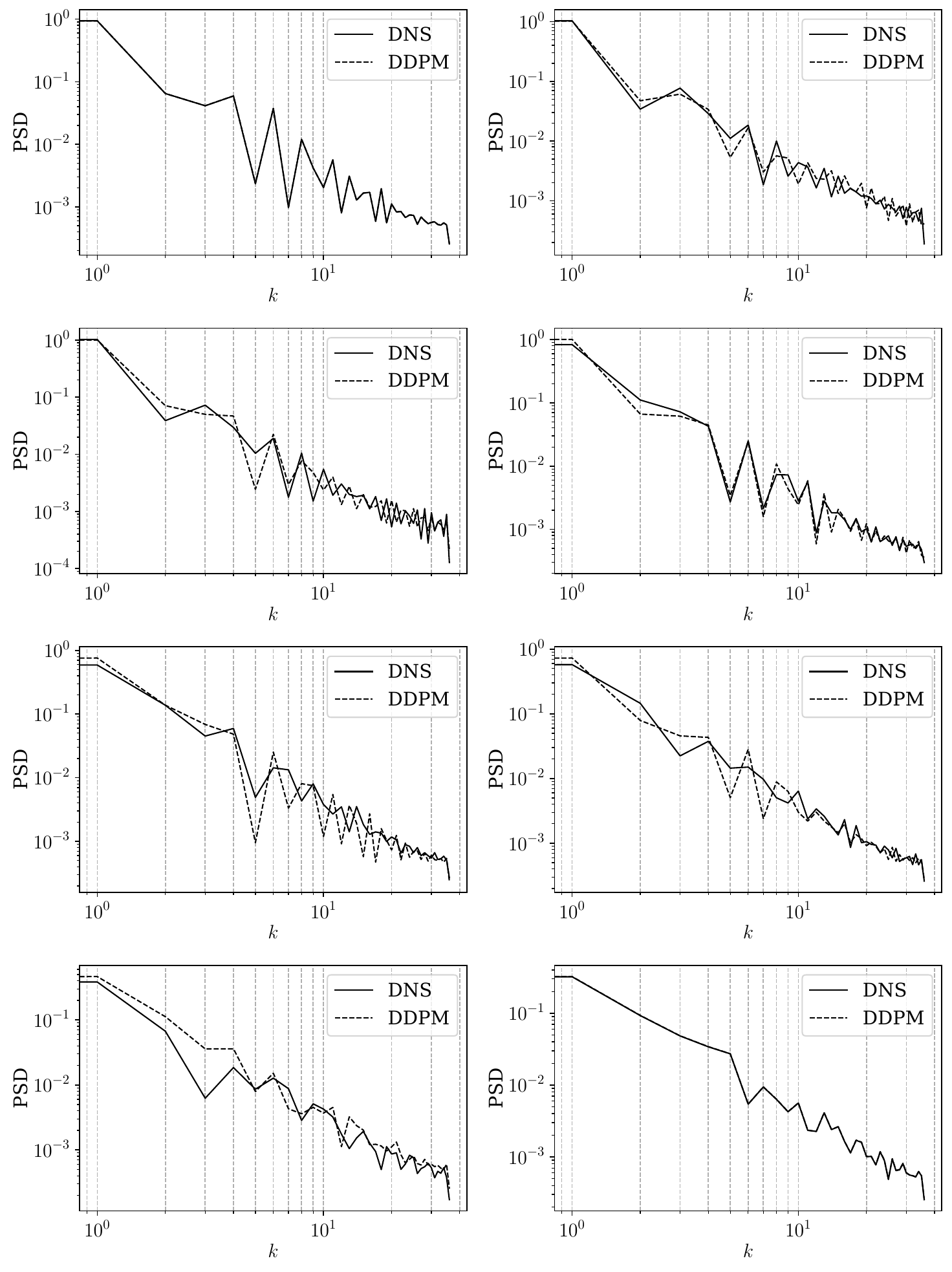}
\caption{Power Spectral Density (PSD) for each of the eight snapshots in the first sequence of the flow, i.e. billow collapse, average in the spanwise direction. Snapshot order is $\{0, 1\}$ in the first row, $\{2, 3\}$ in the second row, $\{4, 5\}$ in the third row, and $\{6, 7\}$ in the last row.} 
\label{fig: kh spectra}
\end{figure}

As the TKE spectrum can be analysed for the velocity field, power spectra are an analogous tool for scalar fields. In this case, we analyse the energy content of temperature fluctuations at different wavenumbers. As before, fluctuations are obtained by removing an average obtained along the spanwise direction. The spectra are then computed for all spanwise locations, and averaged. Figure \ref{fig: kh spectra} details the spectra across frames for the first sequence, i.e. billow collapse. The trend from the ground truth DNS is followed well by the generated sequences in all frames, with the second to last frame showing the largest deviation from the ground truth. 

\begin{figure}[ht!]
\centering
\includegraphics[width=\textwidth, trim=0 100 0 100, clip]{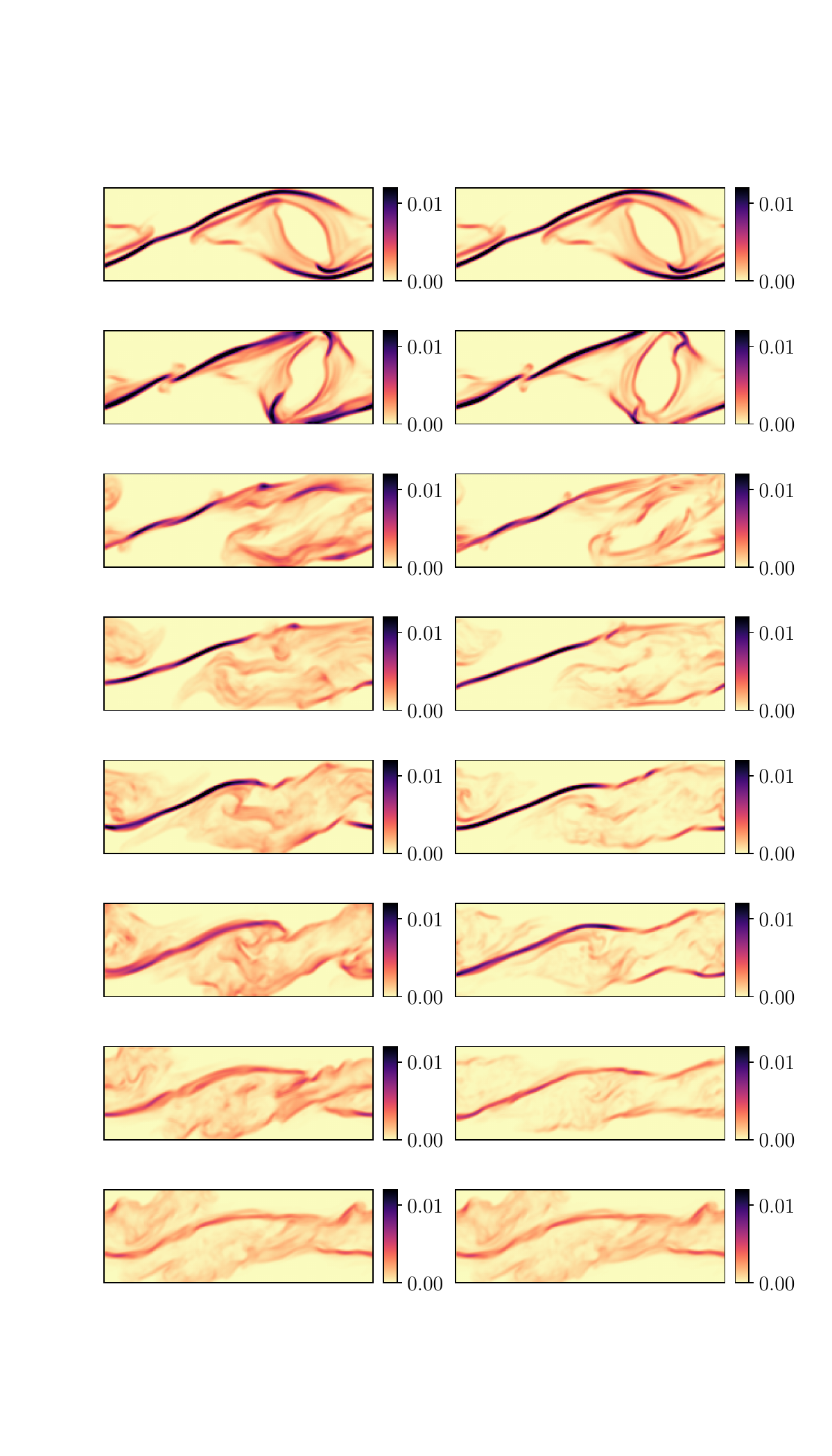}
\caption{Spanwise mean scalar dissipation for the billow collapsing phase of the instability, for ground truth DNS (left), and DDPM generated outputs (right).}
\label{fig:scalar dissipation}
\end{figure}

Scalar dissipation is a constructed parameter which describes the turbulent mixing of a scalar field. In the context of Kelvin-Helmholtz instabilities occurring in thermally stratified flows, this gives an indication of the intensity with which the turbulence is enhancing heat transfer. Accurate recovery of scalar dissipation would entail that the generated flow sequence is capturing the instananeous intensity of thermal energy exchange between the hot flow and cold flow. Figure \ref{fig:scalar dissipation} compares the spanwise mean scalar dissipation field from the ground truth DNS, and the DDPM-generated sequences. Our results demonstrate that the generated sequences are able to statistically capture the dissipation of temperature throughout the flow domain, with some small fluctuations of intensity, and some smaller mixing structures being missed out. 

\section{Conclusions, Considerations, and Future Work}
\label{sec:conclusions}

We have shown that diffusion models may be used to generate statistically correct high-fidelity flow sequences, for both a statistically stationary Kolmogorov flow and a statistically non-stationary Kelvin-Helmholtz instability. We are not aware of prior work which systematically evaluates diffusion-based temporal infilling between sparse, decorrelated turbulent snapshots in the context of a statistically non-stationary instability. 

While the generated outputs from the Kolmogorov flow case were very good, there were some limitations in the sequences generated for the Kelvin-Helmholtz instability. We focused our analyses on the time interval where the statistics change most rapidly, i.e. the billow collapse phase. We found that while the billow collapse itself was recreated well, the turbulence generation as a result of this was captured statistically well, but with minor non-physical artefacts. We hypothesise that the rapidly evolving flow statistics would require smaller $\Delta_t$ between snapshots for this case, and the model would benefit from learning from smoother transitions between phases. 

In order to embed some notion of difference between the different phases, we added class label embeddings to the flow sequences, adding an integer label embedding to the timestep embedding, as a modification of prior work where timestep is concatenated to the input \cite{shi2024}. While this improved results slightly, alternative embedding approaches are yet to be explored. 

Additionally, we observed that with increasing model size, the performance on the Kelvin-Helmholtz instability improved. Thus it may be hypothesised that an even larger model is required; this was beyond our computational resources, but would be worth investigating.

We note that the method introduced in \citet{kim2020} for training a separate model which learns to produce noise for use in a generative method is equally applicable to DDPMs and is worth investigation; the concept of progressive noise has been explored in recent work and shows superior performance over baseline video generation DDPMs \cite{ge2023}. Additionally, there are a number of extensions to video DDPM which have demonstrated strong performance for very long sequences; these are not explored in the current work but would facilitate a true surrogate for DNS precursors. 

\subsection*{Acknowledgements}
The authors would like to acknowledge the assistance given by Research IT and the use of the Computational Shared Facility at The University of Manchester. This work was supported by the Engineering and Physical Sciences Research Council (EPSRC) [EP-T517823-1]. The authors would also like to thank EPSRC for the computational time made available on the UK supercomputing facility ARCHER2 via the UK Turbulence Consortium [EP/R029326/1]. 

\subsection*{Conflict of Interest}
The authors have no conflicts to disclose. 

\subsection*{Data Availability}
Data supporting the findings of this study have been made available at DOI:10.48420/29329565. The supporting code can be found at \hyperlink{https://github.com/HamzaSardar/classifier-free-guidance/releases/tag/Submission2}{https://github.com/HamzaSardar/classifier-free-guidance/releases/tag/Submission2}.

\bibliography{references.bib}
\end{document}